\documentclass[onecolumn,superscriptaddress]{revtex4}

\usepackage[utf8]{inputenc}

\usepackage{graphicx}
\usepackage{amsmath}
\usepackage{slashed}
\usepackage{amssymb}
\usepackage{hyperref}
\usepackage{cleveref}
\usepackage{natbib}
\usepackage[dvipsnames]{xcolor}
\usepackage{tikz}
\usepackage{pgfplots}
\usepackage{xparse}
\usepackage{caption}
\usepackage{subfigure}

\newcommand{\ee}{\end{equation}}
\newcommand{\bb}{\begin{equation}}
\newcommand{\eqb}{\begin{eqnarray}}
\newcommand{\eqf}{\end{eqnarray}}

\def\1{\'{\i}}

\usetikzlibrary{positioning,arrows,patterns,automata}
\usetikzlibrary{decorations}
\usetikzlibrary{trees}
\usetikzlibrary{decorations.pathmorphing}
\usetikzlibrary{decorations.markings}
\usetikzlibrary{calc}

\tikzset{	aphoton/.style={decorate, decoration={snake}, draw=blue},
	photon/.style={decorate, decoration={snake}, draw=black},
	particle/.style={draw=black, postaction={decorate},
		decoration={markings,mark=at position .5 with {\arrow[draw=black]{>}}}},
	aparticle/.style={draw=black},	gluon/.style={decorate, draw=red,
		decoration={coil,amplitude=4pt, segment length=5pt}},
	vertex/.style={draw,shape=circle,fill=black,minimum size=3pt,inner sep=0pt},
}
\NewDocumentCommand\semiloop{O{black}mmmO{}O{above}}
{%
\draw[#1] let \p1 = ($(#3)-(#2)$) in (#3) arc (#4:({#4+180}):({0.5*veclen(\x1,\y1)})node[midway, #6] {#5};)}

\begin{document}
\title{Corrections of $Z'$ to the Magnetic Moment of the Muon}

\author{Jorge Gamboa}
\email{jorge.gamboa@usach.cl }
\affiliation{Departamento de F\'{\i}sica, Universidad de Santiago de Chile, Casilla 307, Santiago, Chile}
\author{Justo L\'opez-Sarri\'on}
\email{justo.lopezsarrion@ub.edu}
\affiliation{Departament de F\1sica Qu\`antica i Astrof\'{\i}sica
and Institut de Ci\`encies del Cosmos (ICCUB), Universitat de Barcelona, Mart\1 Franqu\`es 1, 08028 Barcelona, Spain}
\author{Fernando M\'endez}
\email{fernando.mendez@usach.cl}
\affiliation{Departamento de F\'{\i}sica, Universidad de Santiago de Chile, Casilla 307, Santiago, Chile}
\author{Natalia Tapia-Arellano}
\email{u6044292@utah.edu}
\affiliation{Department of Physics and Astronomy, University of Utah, Salt Lake City, UT 84102, USA}

\date{\today}

\begin{abstract}
We go through several previous corrections and contributions to the muon $g-2$, starting from the dark photon hypothesis to the dark Z. We explore the inputs from a dark Z boson virtual mediator in a first order loop. We consider not only the QED like contributions in the theory but also weak interactions. We obtain a new factor that adds corrections to the form factor associated with the anomalous magnetic moment. We show our result is favorable in new unexplored windows in the mass-coupling parameter space.
\end{abstract}

\maketitle

\section{Introduction}

The search for dark matter has been well established as one of the great challenges of physics today. Dark matter eventual discovery will allow us to understand many puzzles for which the scientific community does not have an explanation based on the currently accepted standard model (SM) of particle physics \cite{RevModPhys.53.1,Kolb:1990vq,goodenough2009possible,PhysRevD.90.015032,DONATO201441,Undagoitia_2015,Baudis_2016,BERTONE2005279,DJOUADI20081}.

{{Interestingly, the search for dark matter allows addressing another fundamental problem of the standard model, namely the puzzle of the gyromagnetic factor of the muon \cite{doi:10.1146/annurev-nucl-031312-120340,miller2007muon,ParticleDataGroup:2016lqr,Aoyama:2020ynm,muoninit} where a  discrepancy between theory and experiment stands to this day and its value is  
\bb
a_\mu^{Exp} -a_\mu^{Th} = (251 \pm 59) \times 10^{-11}, \label{eq:1}
\ee 
suggesting signals of new physics. The results of the Fermilab experiment E989 confirm those found previously at BNL and will be tested once again at the J-PARC experiment in Japan \cite{Iinuma:2011zz}
\footnote{Many efforts have been performed and are currently underway to explain this anomaly through lattice QCD. We do not explore this frontier in this work, but we acknowledge the possibility of groundbreaking results.}. 

In the case of the electron, a big triumph of quantum field theory was  the calculation of the anomalous magnetic moment by Schwinger in 1948 \cite{Schwartz:2014sze,PhysRev.73.416, PhysRev.76.790}. The ``Anomaly" ($a$) refers to the fact that, in contrast to what the Dirac equation says, the Landé factor ($g$) has quantum corrections $a= (g-2)/2 = \alpha / (2\pi)$, to be more specific, the discrepancy described above, which has a prediction by the Standard Model of a relative precision of  $4 \times 10^{-8}$. 
In the case of the electron anomalous magnetic moment, this quantity ($a$) has been calculated up to fifth-loop level \cite{PhysRevD.100.096004}, with the best measurement being 0.25 parts per billion providing an exquisite determination of the fine structure constant, $\alpha$ \cite{anomalousanomaly}.
But, for the muon, any discrepancy with what could be expected from theory opens a window to new physics, which is why so many efforts have been set to measure with better accuracy the muon anomalous magnetic moment ($a_{\mu}$). This difference has been fixed to be $4.2 \sigma$, setting the strongest discrepancy in history for this anomaly \cite{PhysRevLett.126.141801,Baum_2022}.

Different dark matter particles remain among the different candidates that could explain this discrepancy.
The exact mechanism by which the visible and hidden sectors are connected is, of course, currently not known but with some certainty, the relationship between the visible and dark sectors should be dictated by an idea similar to a kinetic mixing  \cite{Holdom1986TwoUA,POSPELOV2000181,PhysRevD.84.103501,PhysRevD.93.103520, Arias_2012,Redondo_2009}. Kinetic mixing is a natural and simple procedure that, after an appropriate Lagrangian diagonalization procedure, allows defining modified vertices that induce processes between visible and dark matter \cite{Fabbrichesi:2020wbt}.

The so called ``dark photon" is a popular possible dark matter candidate for relatively small masses (O(MeV)) and which would give a contribution to the anomalous magnetic  moment, by coupling to the Standard Model through a kinetic mixing, above mentioned \cite{POSPELOV2000181}. This particle is a hypothetical vector boson, and it has been for the most part almost ruled out by different experimental searches within the parameter space of masses which would be favorable in this scenario \cite{Caputo:2021eaa}.

The next natural candidate is what can be found in the literature as the ``Dark Z boson", coupled through kinetic and mass mixing with the electroweak sector of the SM \cite{San:2022uud, PhysRevD.85.115019,ParticleDataGroup:2016lqr}.

A dark Z (DZ) differentiates from the Dark Photon (DP) in its mass mixing with the ordinary Standard Model Z boson and in its weak interactions. Prospect of searches for this new particle has been proposed to be part of the incoming International Linear Collider (ILC), suggesting that with only one month of data, it would be possible to have a measurement of its chiral couplings to fermions with precision of the order of percents \cite{San:2022uud}.

The ideas outlined above concerning a DZ are technically implemented here as follows; instead of the group $U (1)_Y$ of the standard model, we will have the extension $U(1) _Y\times U'(1)$ where $U'(1)$ denotes a gauge boson that we will call $Z'_\mu$ or Dark Z throughout the text. At the Lagrangian level, the general and commonly used kinetic mixing procedure means that instead of the usual kinetic terms 
$F_{\mu \nu} (A) F^{\mu \nu} (A)$ and $F_{\mu \nu} (Z) F^{\mu \nu} (Z)$, the mixing 
$\epsilon F_{\mu \nu} (A) F^{\mu \nu} (Z)$ must be added, where $\epsilon$ is a parameter that should be set by experiments.

However, the presence of the $\epsilon$ parameter --and eventually the mass of the hidden gauge boson--  adds ingredients that lead to non-trivial dynamical modifications, which can be analyzed with data from current experiments.}}

Future experiments previously mentioned that allow limiting the parameters of kinetic mixing will be of central importance in the years to come for the discrepancy between theory and experiment in this case.

Nevertheless, this poses another interesting challenge, that is to say, if $ Z '$ (DZ) physics is a good idea beyond the standard model then it should be able to induce significant corrections at one-loop to improve the agreement between theory and  experiment in the case of the muon's anomalous magnetic moment (for some recent references see for example \cite{PhysRevD.89.095006, alonsoalvarez2021gauging, hapitas2021general, Kayoung, PhysRevD.102.033002}).

We review the previous work of several authors, starting from the DP scenario and continuing with the kinetic mixing set up for the dark $U(1)$ boson ($U(1)_D$); we extend the panorama with a weak kinetic and mass mixing for the DZ \cite{PhysRevD.85.115019} and look for the contributions to the muon anomalous magnetic moment that this framework would convey.

A better understanding of the anomaly $a_{\mu}$ could not only lead to new physics, but it would also be possible to determine if the muon is a composite particle, as it has been theorized, in contrast with its equivalent the electron \cite{anomalousanomaly}.
The most recent $4.2 \sigma$ discrepancy from Fermilab's experiment brings the community closer  to the $5 \sigma$ discovery level required to claim that the Standard Model of particle physics is not able to explain the anomaly, but it is not enough to say with certainty, although this measurement strengthens the evidence for new physics. Future experiments reaching this point doubt the existence of many beyond-the-Standard Model theories, establishing strong limits on them.

A completely different and independent approach to measuring this anomaly remains in the future experiment currently being built at J-PARC in Japan. The main novelty of this experiment is the use of cooled down muons down to a few MeV, which allows them to obtain a high-quality muon beam and enables experiments with much less systematic error than current methods \cite{anomalousanomaly, KESHAVARZI2022115675}.
J-PARC is the only experimental effort with current and future plans to measure the muon g-2, starting engineering and physics runs during 2027 \cite{undefined}.

The paper is organized as follows. In \Cref{sec2} we revisit previous constraints on the dark photon parameter space and show how our method replicates these results. In \Cref{ref:sec3} we extend to a dark boson in general with a $U(1)_Y$ group and compare with literature. \Cref{ref:sec4} shows the contributions coming from electroweak interactions among a dark boson and muons. We discuss our findings and conclude in \Cref{ref:sec5}.

\section{Review of previous results}{\label{sec2}}

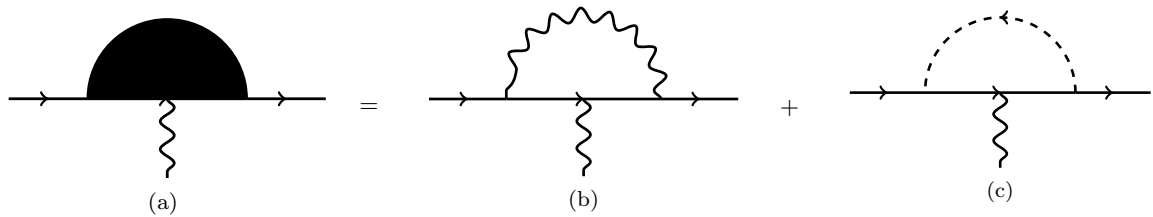
\begin{figure}[t!]
\centering
\begin{minipage}[c]{0.25\textwidth}
  \resizebox{\textwidth}{!}{
\subfigure[\label{fig}] {
\begin{tikzpicture}[node distance=1cm and 1cm]
\coordinate (v);
\coordinate[right=of v] (v1);
\coordinate[right=of v1] (v2);
\coordinate[left=of v](v3);
\coordinate[left=of v3](v4);
\coordinate[below=of v](v5);
\coordinate[below=of v5](v6);
\draw[line width=0.35mm, particle] (v4) -- (v3);
\draw[line width=0.35mm, particle] (v3) -- (v1);
\draw[line width=0.35mm, particle] (v1) -- (v2);
\semiloop[line width=0.35mm, fill=black, aparticle]{v3}{v1}{0};
\draw[line width=0.35mm, photon] (v) -- (v5);
\end{tikzpicture}}}
\end{minipage}
\begin{minipage}[c]{0.05\textwidth}
  $=$
\end{minipage}
\begin{minipage}[c]{0.25\textwidth}
  \resizebox{\textwidth}{!}{
\subfigure [\label{figa}] {
\begin{tikzpicture}[node distance=1cm and 1cm]
\coordinate (v);
\coordinate[right=of v] (v1);
\coordinate[right=of v1] (v2);
\coordinate[left=of v](v3);
\coordinate[left=of v3](v4);
\coordinate[below=of v](v5);
\coordinate[below=of v5](v6);
\draw[line width=0.35mm, particle] (v4) -- (v3);
\draw[line width=0.35mm, particle] (v3) -- (v1);
\draw[line width=0.35mm, particle] (v1) -- (v2);
\semiloop[line width=0.35mm, photon]{v3}{v1}{0};
\draw[line width=0.35mm, photon] (v) -- (v5);
\end{tikzpicture}
}}
\end{minipage}
\begin{minipage}[c]{0.05\textwidth}
$+$
\end{minipage}
\begin{minipage}[c]{0.25\textwidth}
\subfigure [\label{figb}] {
\begin{tikzpicture}[node distance=1cm and 1cm]
\coordinate (v);
\coordinate[right=of v] (v1);
\coordinate[right=of v1] (v2);
\coordinate[left=of v](v3);
\coordinate[left=of v3](v4);
\coordinate[below=of v](v5);
\coordinate[below=of v5](v6);
\draw[line width=0.35mm, particle] (v4) -- (v3);
\draw[line width=0.35mm, particle] (v3) -- (v1);
\draw[line width=0.35mm, particle] (v1) -- (v2);
\semiloop[line width=0.35mm, dashed, particle]{v3}{v1}{0};
\draw[line width=0.35mm, photon] (v) -- (v5);
\end{tikzpicture}
}
\end{minipage}
\captionsetup{font=footnotesize}
\captionof{figure}{\label{fig1}The two possible vertex corrections involved 
in the gyromagnetic factor of dark matter (dashed lines represent $Z'$ 
 propagators).  Diagram (a) is the total anomalous magnetic moment contribution to the muon vertex. This diagram represents the sum of all possible contributions, here we list only two. Diagram (b) is the QED type contribution from a dark vector boson, the dark photon. The last diagram (c) shows the contribution of a weak vertex contribution from a Z' boson. 
}
\end{figure}

A minimal extension of the Standard Model is a $U(1)$ group, coupled through a kinetic mixing to the visible photon. There is a broad amount of publications explaining the detail of this mechanism, and examples can be found vastly in the literature. We suggest some examples in the next papers \cite{PhysRevD.80.095002, Holdom1986TwoUA,POSPELOV2000181,PhysRevD.84.103501,PhysRevD.93.103520, Arias_2012,Redondo_2009}.

To fix the notation we follow our previously used prescription  \cite{Das:2016cyx,Schwartz:2014sze}.
We identify as $q^\mu_1$ and $q^\mu_2$ the incoming and outgoing momenta,  then the transferred momentum would be $p^\mu = q^\mu_1 -q^\mu_2 $, according to what we have set and shown in \cref{fig1}. Under such conditions, the structure of the vertex function has the form 
\bb
i{\cal M}^\mu = -\,i  e {\bar u}(q_2) \Gamma^\mu u(q_1), \label{gene}
\ee
with $\Gamma^\mu$ function given by 
\bb
\Gamma^\mu = \gamma^\mu F_1 (\frac{p^2}{m_\mu^2}) + \frac{\sigma^{\mu \nu}}{2m_\mu} p_\nu F_2(\frac{p^2}{m_\mu^2}), \label{vertex}
\ee 
where $F_1$ and $F_2$ are two Lorentz invariant form factors.

If we draw our attention to the third order interaction term in the electric charge  --for example the diagram in 
\Cref{figa}, with a photon in the loop--  the scattering amplitude becomes

\bb
i{\cal M}_b^\mu = -\,e^3 \epsilon^2 {\bar u}(q_2) \int \frac{d^4k}{(2\pi)^4} 
\frac{\eta_{\alpha\beta} 
\gamma^\alpha ({{  p\hspace{-.6em}  \slash \hspace{.15em}}} +
{{ k \hspace{-.6em}  \slash \hspace{.15em}}} + m_\mu) 
\gamma^\mu 
({{ k\hspace{-.6em}  \slash \hspace{.15em} + m_\mu)
\gamma^\beta}}}{[(k-q_1)^2 + i\varepsilon] 
[(p+k)^2 - m_\mu^2 + i\varepsilon] [k^2 -m_\mu^2] + i\varepsilon]} \label{diaa}
u(q_1), 
\ee
\noindent 

while for \Cref{figb}, where the loop carries a massive boson, we have 

\bb
i{\cal M}_c^\mu = -\, e^3  {\bar u}(q_2) \int \frac{d^4k}{(2\pi)^4} 
\frac{\eta_{\alpha\beta} 
\gamma^\alpha ({{  p\hspace{-.6em}  \slash \hspace{.15em}}} +
{{ k \hspace{-.6em}  \slash \hspace{.15em}}} + m_\mu) 
\gamma^\mu 
({{ k\hspace{-.6em}  \slash \hspace{.15em} + m_\mu)
\gamma^\beta}}}{[(k-q_1)^2 -m_Z^2+ i\varepsilon] 
[(p+k)^2 - m_\mu^2 + i\varepsilon] [k^2 -m_\mu^2] + i\varepsilon]} \label{diab}
u(q_1),
\ee

\noindent where $m_Z$ represents the mass of the boson in question, independently of it being a dark photon, a Z boson or a Dark Z.

The interesting question is how to calculate the form factors, with the purpose of obtaining our desired contributions to $a_{\mu}$. To do this, we first note that the physical interpretation of $F_1$ and $F_2$ is clear, $F_1$ leads to  charge renormalization while $ F_2 $ is a direct contribution to the magnetic moment.

Taking the above into account, we will concentrate on properly isolating each diagram corresponding to  (\cref{diaa}) and (\cref{diab}) so we can identify the factors proportional to $\sigma^{\mu \nu}$. 

If we use Gordon's identity, after some algebra we find

\bb 
F^{(c)}_2 (p^2) = - i\,8\,m_\mu^2 e^2 \epsilon^2 \int_0^1 dx dy dz \delta (x+y+z-1) \int \frac{d^4k}{(2\pi)^4} \frac{z(1-z)}{(k^2-\Delta + i\varepsilon)^3} + \cdots 
\ee
with $\Delta = -x y p^2 + (1-z)^2 m_\mu^2 + m_Z^2 z$ and $x,y$ and $z$, Feynman parameters \cite{Schwartz:2014sze}.

Once $F_2 (p ^ 2)$ is given, then we evaluate it at the limit $p \to 0$ and we obtain 

\bb
a_\mu= \frac{g_\mu -2}{2} = F_2(0).
\ee

In the limit $m_Z \to 0$ we obtain that $F_2 (0) = \frac{\alpha}{2 \pi}$ according to the classical Schwinger result for the massless SM photon \cite{PhysRev.73.416, Schwartz:2014sze}. 

If $ m_Z \neq 0$, the result is

\begin{equation}
a_{\mu}=  \frac{\alpha}{2 \pi} (\epsilon^2)  f(\kappa),
\end{equation}

\noindent with $f(\kappa)$, coming from the above integration, defined as

\begin{equation}
f(\kappa)=
   \left\{1-2 \kappa ^2+2 \left(\kappa ^2-2\right) \kappa ^2 \log (\kappa )-\frac{ \left(\kappa ^4-4 \kappa ^2+2\right) \kappa  \tan^{-1} \left(\left(\sqrt{4-\kappa ^2}/\kappa \right)\right)}{\sqrt{4-\kappa ^2}}\right\}.
   \label{eq:f_nog5}
\end{equation}

\begin{figure}[t]
    \centering \captionsetup{justification=centering} \includegraphics[scale=0.75]{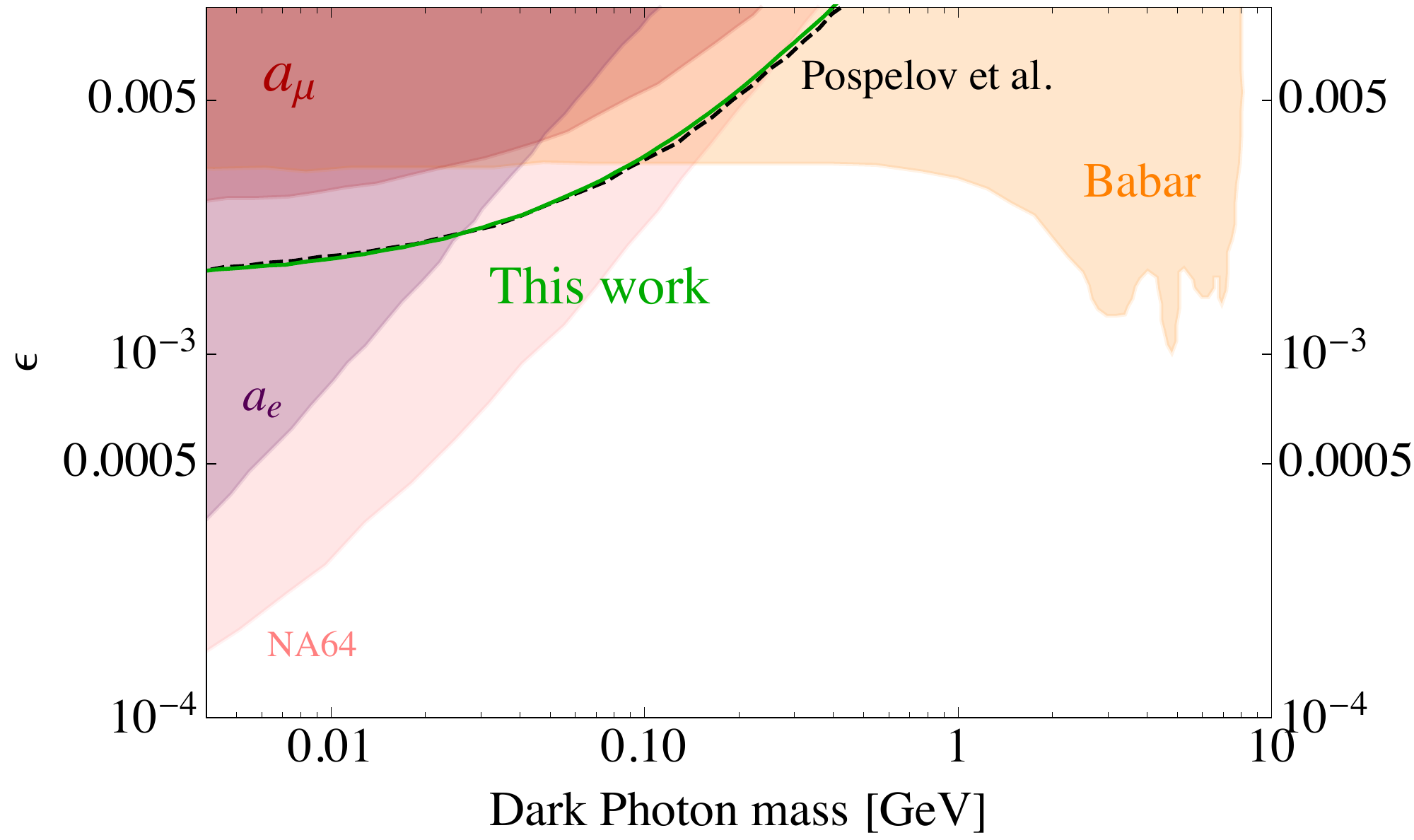}
    \caption{In green, our analytic expression in \Cref{eq:f_nog5} for the numerical results previously outlined by Pospelov et al. (see \cite{PhysRevD.80.095002}) marked in black dashed lines.
    This is the case of a $U(1)$ sector coupled to photons.
    For comparison we show known restricted zones from Babar \cite{thebabarcollaboration2008search,PhysRevD.91.094026} and NA64 experiments \cite{PhysRevLett.118.011802}, along with SM results for the exclusion in parameter space to the electron and muon anomalous magnetic moment. Constraints adapted from \cite{PhysRevLett.118.011802}.
    }
    \label{fig:pospelov}
\end{figure}

\begin{figure}[t]
    \centering    \includegraphics[scale=0.75]{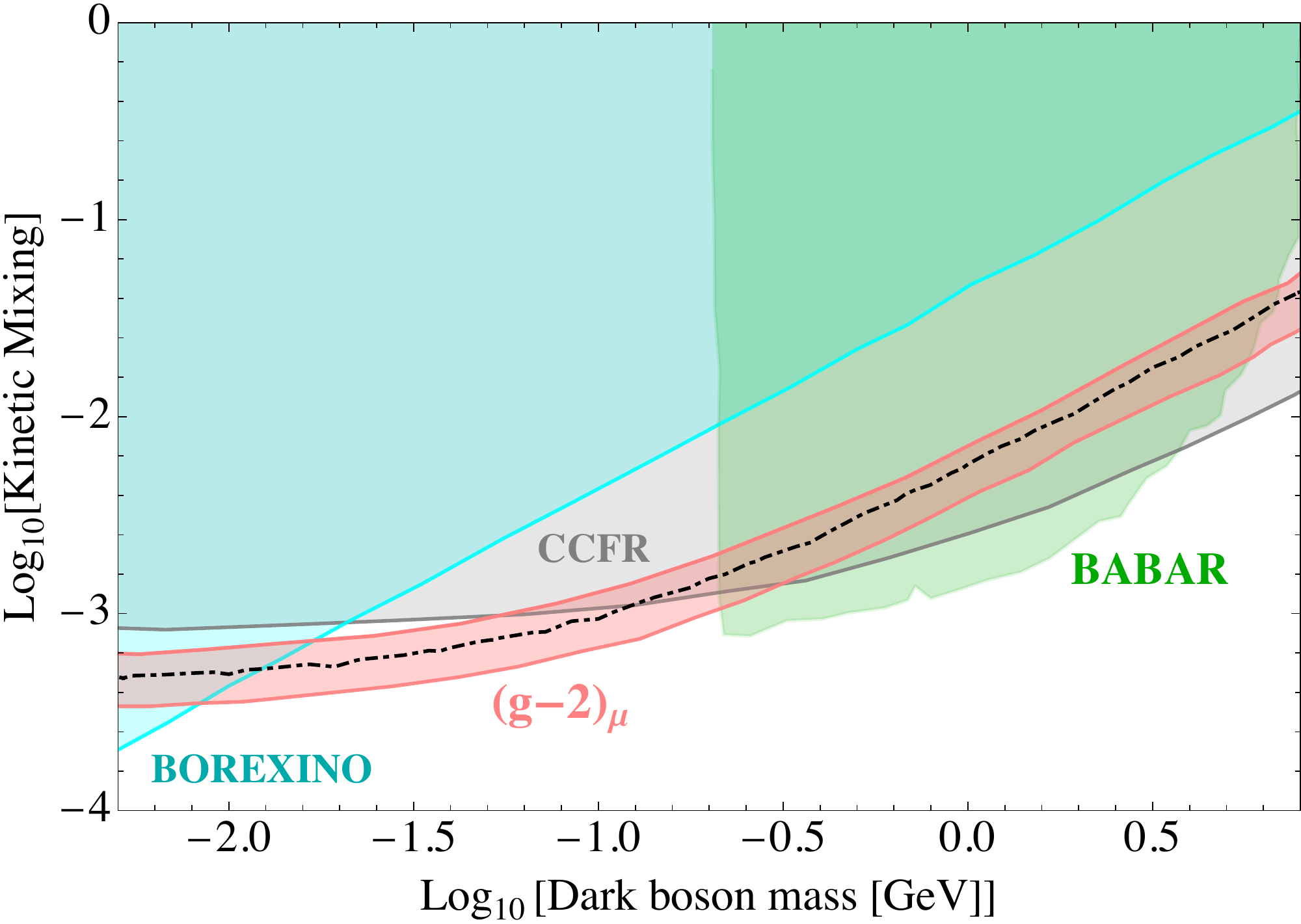}
    \caption{Our analytic expression in \ref{eq:f_nog5} is shown in black dashed lines, this is the U(1) coupling to SM.
    Constraints on the Dark Boson mass and kinetic mixing from \cite{PhysRevD.103.015008} can be seen as shaded regions, plus their result for the muon $g-2$ using this symmetry, again (as in \cref{fig:pospelov}) our result explains in a complete and legible expression the previous result fitted to the anomaly. Babar excluded region comes from searches of $Z'$ or Dark Z from the productions of $\mu^{-} \mu^{+} Z'$ at colliders \cite{PhysRevD.94.011102}. CCFR comes from  measurement of the neutrino trident cross section \cite{PhysRevLett.113.091801} and Borexino from neutrino electron scattering \cite{PhysRevD.98.015005}.
    In \Cref{fig:final} we use these constraints, but there we include a mass mixing and a hypercharge coupling to the SM. Figure adapted from \cite{PhysRevD.103.015008}. }  
    \label{fig:MZ_us_nog5}
\end{figure}

This analytic expression is equivalent to that in \cite{PhysRevD.80.095002} for \cref{diaa} and we can reproduce those results as can be seen in \Cref{fig:pospelov} in the case of a dark photon and 
in \Cref{fig:MZ_us_nog5} for a dark Z. Both Figures show the contribution from our analytical expression in \Cref{eq:f_nog5}. We take this point as the motivation for our work since the dark photon hypothesis favoured parameter space has been severely tested and not found, as shown in \Cref{fig:pospelov}. We seek new unexplored regions that can be enlightened by searches related to a Dark Z boson. Furthermore, \Cref{fig:MZ_us_nog5} shows that the same mechanism to obtain constraints for the case of the DZ has a small parameter space allowed for future searches. 

It is important to highlight that this correction has been obtained assuming that only QED-like couplings contribute to the $\Delta a_{\mu}$ in the case of a massive virtual boson in the loop. 
{A similar approach as been taken on Arcadi et. al. \cite{ARCADI2022115882}}

In the next section, we explore the contributions of electroweak-like interactions.

\section{Consequences of dark electroweak interactions with Z' boson}{\label{ref:sec3}}

In the standard model, the kinetic terms of the Lagrangian after electroweak symmetry breaking are:
\begin{equation}
    \mathcal{L}_{kin}=-\frac{1}{4} F_{\mu \nu}^2 -\frac{1}{4} Z_{\mu \nu}^2,
\end{equation}
and the neutral currents coupled to the Z and photon bosons can be written as:
\begin{equation}
    \mathcal{L}=\mathcal{L}_{kin} + \frac{e}{\sin \theta_{W}} Z_{\mu} J_{\mu}^Z + e A_{\mu} J_{\mu}^{EM},
    \label{eq:kineplus}
\end{equation}
where we should keep in mind that here that $e=g \sin \theta_W = g' \cos \theta_W$, comes from the covariant derivative $D_{\mu} H= \partial_\mu H -i g W_{\mu}^a \tau^a H - 1/2 i g' B_{\mu} H$ with $W_{\mu}^a$ and $B_{\mu}$ the SU(2) and $U(1)_Y$ gauge bosons, respectively and $\theta_{W}$ the usual Weinberg angle.
The explicit currents, $J_{\mu}^Z$ and $J_{\mu}^{EM}$ are
\begin{eqnarray}
    J_{\mu}^Z & = &\frac{1}{\cos\theta_W} \left( J_{\mu}^3- \sin^2 \theta_W J_{\mu}^{EM} \right),
\\
    J_{\mu}^3 & = & \sum_i \bar{\psi}_i^L \gamma_\mu T^3 \psi_i^L, \quad J_{\mu}^{EM} = \sum_i (T^3 + Y)\bar{\psi}_i^L \gamma_\mu  \psi_i^L.
\end{eqnarray}

The Lagrangian including neutral current interactions, \Cref{eq:kineplus}, can also be expressed as:

\begin{equation}
    \mathcal{L}=\mathcal{L}_{kin}  +
    \frac{g}{\cos \theta_W} \bar{\Psi} \gamma^{\mu} \left[C_L \frac{1-\gamma^5}{2}+
    C_R \frac{1+\gamma^5}{2}\right]\Psi+ e A_{\mu} J_{\mu}^{EM},
\end{equation}

\noindent where

\begin{equation}
    C_L=T^3-Q\sin^2 \theta_W \quad C_R=-Q \sin^2\theta_W.
\end{equation}

If now we mix $B_\mu$ and $Z'_\mu$ which live in $U(1)_Y$ and $U(1)'$ associated with the hypercharge and dark gauge symmetry respectively, then the dynamics of the two gauge fields interacting through a kinetic mixing is
 
\bb
{\cal L} \supset -\frac{1}{4} F_{\mu \nu} (B)F^{\mu \nu} (B)  -\frac{1}{4} F_{\mu \nu} (Z')F^{\mu \nu} (Z') +  \frac{\epsilon}{2 \cos \theta_W} F_{\mu \nu} (B) F^{\mu \nu} (Z'), \label{2}
\ee 
where 
\bb
F_{\mu \nu} (B) =\partial_\mu B_\nu - \partial_\nu B_\mu, ~~~~F_{\mu \nu} (Z') =\partial_\mu Z'_\nu - \partial_\nu Z'_\mu,  \nonumber
\ee 
and $\epsilon$ is a small dimensionless parameter, usually denoted as a kinetic mixing term.

The Lagrangian \ref{2} can be diagonalized by performing the transformation $B' = B + \epsilon/ \cos \theta_W Z'$ from which we obtain 

\bb
{\cal L} \supset -\frac{1}{4} F_{\mu \nu} (B)F^{\mu \nu} (B)  -\frac{1}{4} \left(1-\frac{\epsilon^2}{\cos \theta_W^2} \right) F_{\mu \nu} (Z')F^{\mu \nu} (Z') , \label{3}
\ee 
so the only effect of kinetic mixing is to redefine the charge.

If we include fermion fields (muons), the Lagrangian will now be

\bb
\left(-e \varepsilon J_{em}^\mu  
+  \frac{g}{\cos \theta_W } \varepsilon_Z J_{\mu}^Z 
\right)
Z^{\mu '} , 
\label{4}
\ee
\noindent 
where we see that the diagonalization of the Lagrangian (\ref{3}) modifies the couplings, giving as a result the next interaction term:

\begin{equation}
    \mathcal{L} \supset \frac{-g}{2 \cos \theta_W} \varepsilon_Z J_{\mu}^{Z '} Z^{\mu '}  
\end{equation}

As an important observation, we point out that the inclusion of kinetic mixing induces a non-trivial redefinition of the parameters of the responsible theory, on the one hand, for the appearance of millicharges $\epsilon e$ and on the other, for the need to adjust according to experimental data the re-definitions of the effective coupling constants.

The above expression resembles the QED Lagrangian in calculating the muon magnetic moment for the relevant diagrams corresponding to \Cref{fig1}. These diagrams contain the conventional electromagnetic interaction first while the second is due to the interaction with a massive boson, in our case the $Z'$ boson. Furthermore, the above is the  interaction term for a $m_Z$ boson with photon-like coupling to the fermion. 
Suppose we consider the electroweak corrections to the g-2 regarding a $Z$. In that case, our result can be verified following the Particle Data Group results for electroweak contributions. Still, in this case, we have to re-arrange terms depending on the unknown mass of this new particle \cite{Workman:2022ynf,1864727}.

The kinetic mixing is generalized by including a mass mixing between the standard model $Z$ and the $Z'$ with the introduction of the mass matrix

\begin{equation}
 M_0^2=m_Z^2
 \begin{pmatrix}
1 & -\varepsilon_Z \\
-\varepsilon_Z & m^2_{Z'}/m_Z^2
\end{pmatrix},
\end{equation}

\noindent with the mixing parametrized by

\begin{equation}
    \varepsilon_Z=\frac{m_{Z'}}{m_Z}\delta.
\end{equation}

The degree of mass mixing between the Standard Model Z and the dark Z is given by the $\delta$ parameter above. The former conveys a decay channel for the Higgs of the form $H\rightarrow Z Z'$ which allows to set constraints from atomic parity violation, polarized $e$ scattering and rare $K$ and $B$ decay. We initially set bounds in parameter space using values from Davoudiasl et. al.\cite{PhysRevD.85.115019}, but since their work and the experimental discovery of the Higgs boson, $\delta$ has been further constrained.

\section{The Z' correction}{\label{ref:sec4}}

{In the case of electroweak corrections we have gone through the amplitude calculations with the {\tt Feyncalc} and {\tt FenyArts} {\tt Mathematica} package\cite{Hahn:2000kx} to replicate the PDG's result on the contributions to the muon anomalous magnetic moment. We can recover the terms coming from loops involving Higgs, Z boson and neutrinos (or any of the weak bosons). }

{To procure these contributions we have followed the steps of \cite{PhysRevD.6.2923,PhysRevD.5.2396,PhysRevD.6.374} and as a result we get the  amplitude}

\begin{equation}
\begin{split}
i{\cal M}_c^\mu = -\, {\bar u}(p') \int \frac{d^4k}{(2\pi)^4} &
\frac{1}{((k+p)^2 - m_{\mu}^2)\cdot((k+p')^2 - m_{\mu}^2)\cdot(k^{2}-m_{Z}^2)}~\times \\
 &\Bigg\{ \left(-ie\frac{ \gamma^\nu (1-\gamma^5) (\sin \theta_W^2-1/2) }{2 \sin \theta_W \cos \theta_W}
-ie\frac{ \gamma^\nu (1+\gamma^5) \sin \theta_W }{2 \cos \theta_W}\right)\times  \\
 &((\slashed k+\slashed p')+m_{\mu})\cdot\gamma^{\mu}\cdot((\slashed k+\slashed p)+m_{\mu})~\times \\
 &\left(-ie\frac{ \gamma^\nu (1-\gamma^5) (\sin \theta_W^2-1/2) }{2 \sin \theta_W \cos \theta_W}
-ie\frac{ \gamma^\nu (1+\gamma^5) \sin \theta_W }{2 \cos \theta_W}\right)\Bigg\} u(p).
\end{split}
\label{eq:amplitudeZ}
\end{equation}

From this result, we can obtain the form factors, as was done before in the case of a dark photon, the new interaction terms will give us new corrections to the vertex for a dark Z, but before that, we will shortly review the appropriate Gordon Identity in the following section \ref{sec:gordon}.

\subsection{The Gordon Identity for weak contribution
\label{sec:gordon}}

We use the Pauli form factor defined by the most general form of current conservation and CP invariance \cite{PhysRevD.6.374}, which differs from the usually seen on the QED vertex corrections.

The modified vertex has the form:

\begin{equation}
    \bar{u}(p+q)\Lambda_{\mu} u(p)=-ie \bar{u}(p+q) \left( \gamma_{\mu} F_{1} (q^2) + \frac{i}{2 m_{\mu}} \sigma_{\mu\nu} q^{\nu} F_{2} (q^2) + (q^2 \gamma_\mu -q \cdot\gamma q_{\mu} ) \gamma_{5} F_{3}(q^2) \right) u(p).
\label{eq:Gordon_weak}
\end{equation}

The contribution to the form factors can be  arranged in terms of the factors proportional to $\gamma_\mu$, $\sigma_{\mu\nu} q^{\nu}$ and $\mathbf{\gamma_{5}}$\cite{PhysRevD.6.374}. From \cref{eq:amplitudeZ} we can see that several terms would contribute to $F_3$ but, after some algebra, some terms will take the shape of our desired form factor $F_2$, which will contribute to the correction of the magnetic momentum.

{
We start by selecting the Feynman diagrams associated with \ref{fig1}, and we work with the amplitude shown in \ref{eq:amplitudeZ}. To the resulting expression,  we apply the Gordon Identity as described in \cref{eq:Gordon_weak} and we identify the corresponding contribution to the form factor $F_2$.}

We expect to obtain a contribution proportional to the one in the Standard Model correction of $a_{\mu}$, since the propagators have a similar shape and the main differences are the masses and constants. This is an important contrast respect  to our previous work \cite{Das:2016cyx}; in the case of a dark-QED we go from a massless photon to a massive one and the propagators have a different form.

With the help of {\tt FeynCalc}, we obtain the correspondent form factor $F_{2}$ in terms of Passarino-Veltman integrals, here shown as the different $C_i$ coefficients:

\begin{equation}
    i{\cal M}_c^\mu = C_{0} + 2\, C_{1} + \left( 8\, \sin^4\theta_W - 4\, \sin^2 \theta_W +1 \right) \left( C_1+C_{11}+C_{12} \right)
    \propto F_{2},
\label{eq:PAVE}
\end{equation}

\noindent where $C_0,\, C_1\,, C_{11}\, \text{and}\, C_{12}$ are the integrals depending on combinations of the masses involved \cite{Patel:2015tea}, that is, the muon and $Z$ in the present  case. 

We rapidly check the Standard Model result using {\tt Package X} \cite{Patel:2015tea} on these integrals. By approximating and expanding in a series on the ratio of the masses $m_{\mu}/m_{Z}$  we recover the result below, for contributions of weak interactions to the $F_{2}$, as can be seen on \cite{ParticleDataGroup:2016lqr}:

\begin{equation}
    a_{\mu} \propto \frac{\pi ^2}{\sqrt{2}} G_W m_{\mu }^2 \frac{1}{3}\left(5-\left(1-4\,   \sin^2 \theta_W \right)^2\right).
\end{equation}

To take all the contributions adding up to the $\Delta a_{\mu}$ from dark matter in \cref{eq:PAVE}, we define the ratio $\tau = M_{Z'}/m_{\mu}$ and each component takes the form:

\begin{figure}[b!]
    \centering
    \includegraphics[scale=0.43]{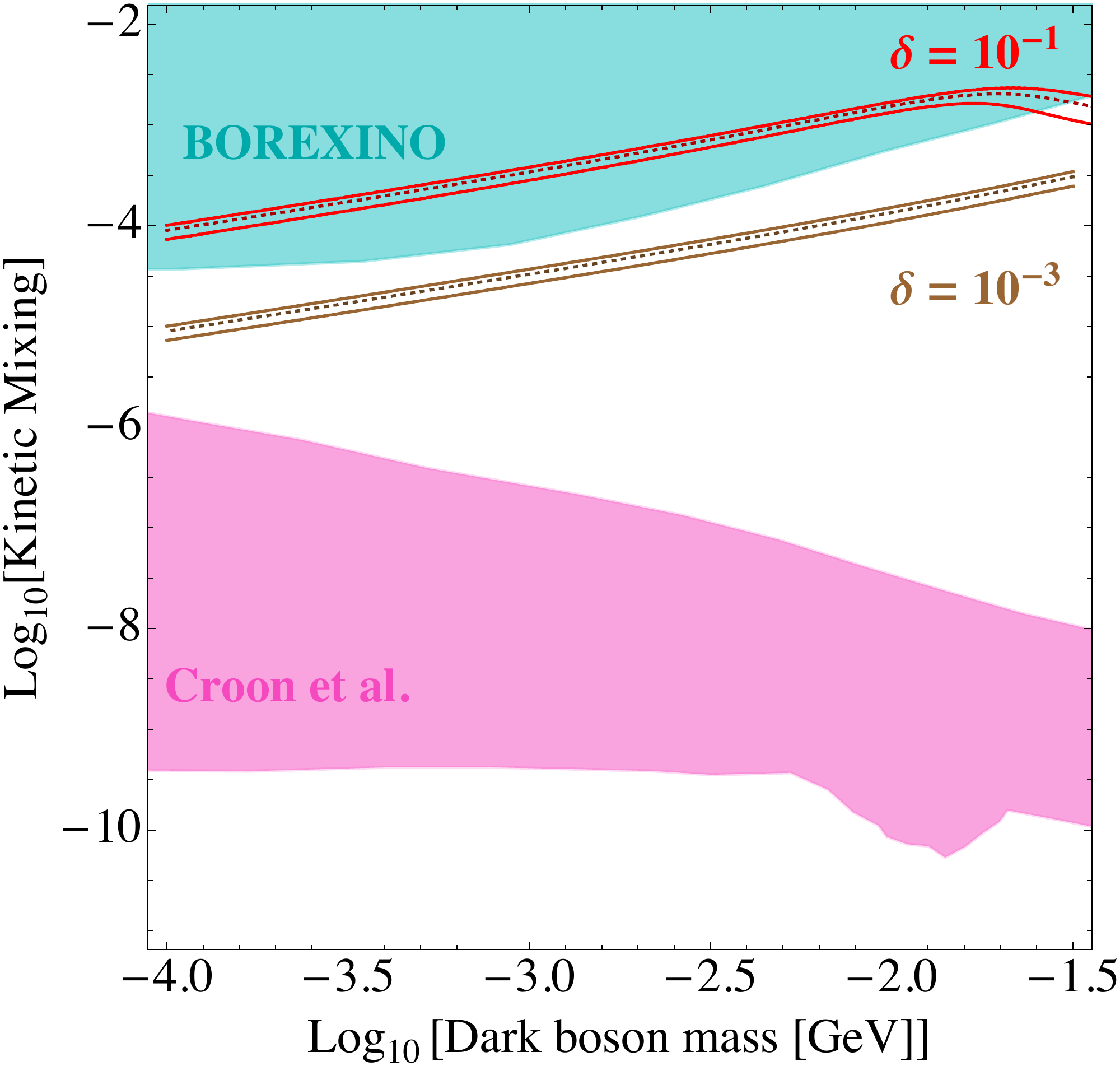}
    \includegraphics[scale=0.425]{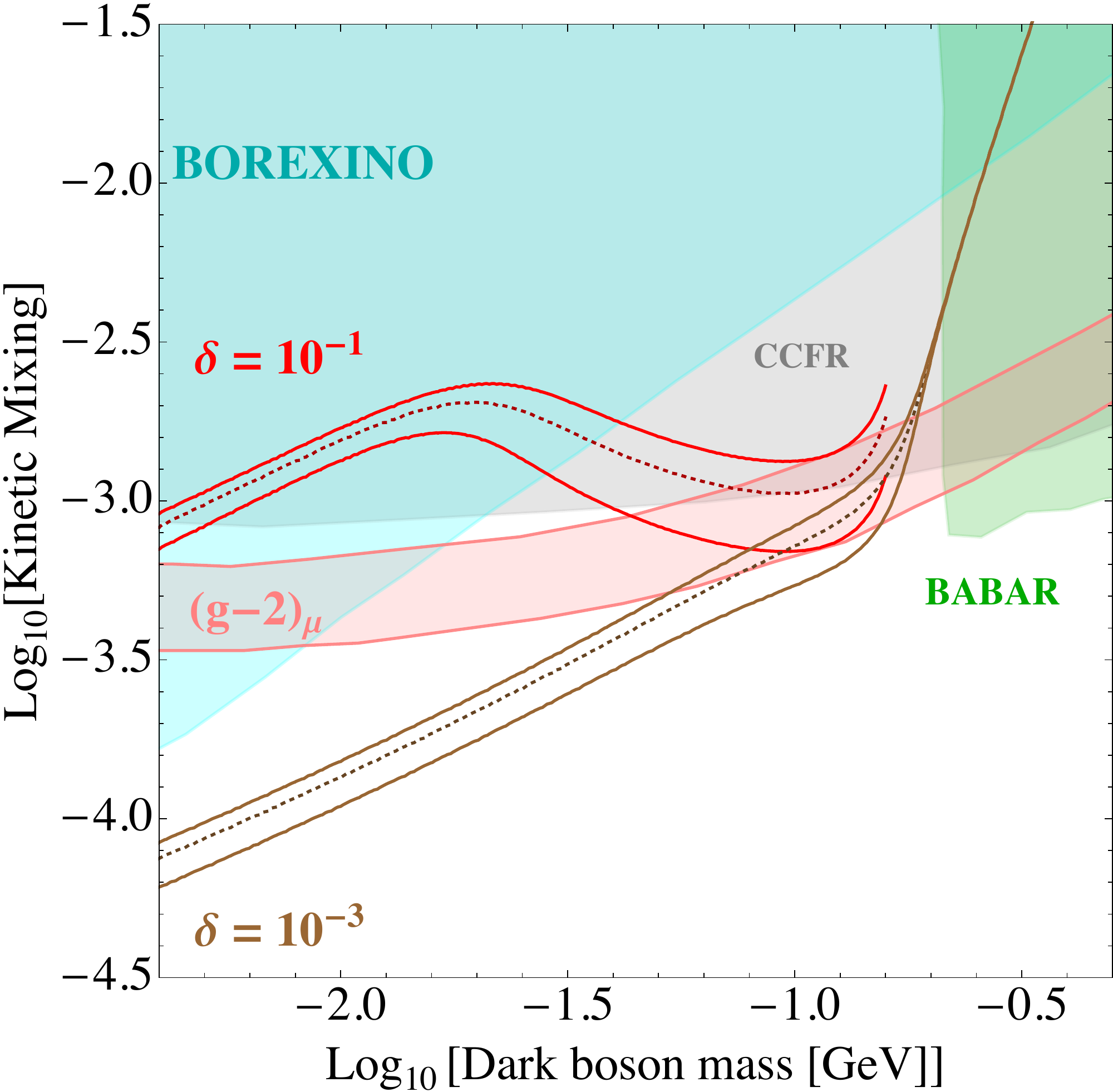}
    \caption{Constraints for the Dark Z boson (DZ) in the kinetic mixing versus boson mass parameter space in two regimes of the mass mixing parameter, $\delta$, from \cref{eq:ammuZ}.
    We set $ \delta $ to be $10^{-3}$ and $10^{-1}$, shown as the brown and red contours, respectively. 
    In a dotted line, we represent the exact comparison between our result and the anomaly, \cref{eq:1}. The straight lines by the sides of the dotted one represent the $2\sigma$ allowed region.
    {\textsl{Left:} In fuchsia, constraints outlined by Croon et al.\cite{Croon:2020lrf} on supernova muons coupled to Z' and Borexino, from \Cref{fig:MZ_us_nog5}}. {\textsl{Right:} In pink, exclusion zone from previous work in dark photon-QED like approach to a DZ \cite{PhysRevD.103.015008} as shown in \cref{fig:MZ_us_nog5} and coming from simply setting the masses for the boson in \cref{eq:f_nog5}, plus the constraints from Babar, Borexino and CCFR as described in the \Cref{ref:sec5}.
    }}
    \label{fig:final}
\end{figure}

\begin{equation}
\begin{split}
a_{\mu}(Z)&= 
\frac{\alpha}{2 \pi} (\varepsilon^2)
\left\{
\bar{f}(\tau)
+ \left[ 8\, \sin^4 \theta_{W} ' -4\, \sin^2 \theta_W '  +1\right]  f(\tau)
\right\}
\end{split}
\label{eq:ammuZ},
\end{equation}

\noindent with $f(\kappa)$ given by \cref{eq:f_nog5},  $\bar{f}(\tau):$

\begin{equation}
\bar{ f}(\tau)
=
-\frac{\tan^{-1} \left(\frac{(\sqrt{4-\tau^2})}{\tau^2}\right)}{\sqrt{4-\tau^2}} + 1 - (\tau^2-1) \log(\tau)
\end{equation}

\noindent and, according to \cite{PhysRevD.85.115019}

\begin{equation}
\sin^2 \theta_{W} ' = \sin^2 \theta_{W}- \frac{\varepsilon}{\varepsilon_Z}  \sin \theta_{W} \cos \theta_{W},
\end{equation}

\noindent such that, by taking the definition of $\varepsilon_Z= \frac{m_{Z'}}{m_Z}\delta $ for simplicity, we will get

\begin{equation}
    \sin^2 \theta_{W} ' = \sin^2 \theta_{W}- \frac{m_{Z} \varepsilon}{m_Z'  \delta}   \sin \theta_{W} \cos \theta_{W}.
\end{equation}

We use \cref{eq:ammuZ} to be compared with the expected value from the SM for $a_{\mu}$ and show it in the parameter space of kinetic mixing versus dark boson mass.
Our final results are shown in \Cref{fig:final}.

\section{Results \&
 discussion}{\label{ref:sec5}

To set the constraints shown in \Cref{fig:final}, for these two parameter space windows, we vary the kinetic mixing $\varepsilon$ and the dark boson mass, the DZ, using our result in \ref{eq:ammuZ}. Consequently, since we have a mass mixing, we set the value of $\delta$ for the mass matrix. We have fixed $\delta$ to span the two extremes of parameters described in \cite{PhysRevD.85.115019}, being $10^{-3}$ and $10^{-1}$, shown in \Cref{fig:final} as the brown and red contours, respectively. 
These come from atomic low energy parity violating experiments, polarized electron scattering, and bounds on the mixing obtained from rare K and B decays.
The dashed lines in \Cref{fig:final} represent the correspondence between our result, in \Cref{eq:ammuZ} to the most recent, $4\sigma$ experimental discrepancy, in \cref{eq:1}. This means that we adjust our result to the \textit{whole} anomaly. The solid curved lines, forming a band, represent the $2 \sigma$ deviation from that result in each $\delta$ extreme described above, following results previously outlined in \cite{PhysRevD.85.115019}. Our analytical expression and results allow us to set discovery lines for a longer range of parameters. 

We focus on two regions of interest, with previously restricted regions set up by Croon et al. \cite{Croon:2020lrf} in pink and Borexino, represented in both windows of \Cref{fig:final} in cyan shades. The right panel of \Cref{fig:final} shows CCFR and Babar constraints from  the measurement of the neutrino trident cross section \cite{PhysRevLett.113.091801} and from neutrino electron scattering \cite{PhysRevD.98.015005}, respectively, besides the favoured zone by 
\cite{PhysRevD.103.015008} regarding their results on $a_{\mu}$.

We have set new windows for the search of a hypothetical dark matter candidate, the dark Z. This prospective dark matter particle still interacts with the SM through kinetic mixing but through a
hypercharge, as well as mass mixing.

The anomalously behaving muon is a lucky case to investigate and probe the limits of our knowledge. Its anomaly is much more interesting than the electron's anomaly, being short lived and unstable and much more sensitive to hypothetical physics beyond the SM, compared with its counterpart, the $\tau$, for which we do not have the technology even to measure a possible anomaly. The possibility of new physics coming from the $\Delta a_{\mu}$ has kept the scientific community working on its explanation for years, and the fact that its equivalent, the electron, has a so well understood and precise calculation and measurement makes us think that this possibility is a reality. A chance to understand any physics beyond the Standard Model predictions opens a window to connect with other phenomena and mysteries, like the dark matter particle composition itself.

\section{Acknowledgments}

We would like to thank  Prof. J. L. Cort\'es for discussions in the early stages of this paper. 
This research was supported by DICYT 042131GR and Fondecyt 1221463 (J.G.), DICYT 042231MF (F.M.), JLS thanks the Spanish Ministery of Universities and the European Union Next Generation EU/PRTR for the funds through the Maria Zambrano grant to attract international talent 2021 program.
The work of NTA is supported by the University of Utah.

\bibliographystyle{unsrt}
\bibliography{references}
\end{document}